\documentclass[journal=jacsat,manuscript=article]{achemso}

\usepackage[version=3]{mhchem}
\usepackage{color}

\author{Klichchupong Dabsamut}
\affiliation{Department of Physics, Faculty of Science, Kasetsart University, Chatuchak, Bangkok 10900, Thailand}
\author{Thanasee Thanasarnsurapong}
\affiliation{Department of Physics, Faculty of Science, Kasetsart University, Chatuchak, Bangkok 10900, Thailand}
\author{Intuon Chatratin}
\affiliation{Department of Materials Science and Engineering, University of Delaware, Newark, Delaware 19716, USA}
\author{Tosapol Maluangnont}
\affiliation{Electroceramics Research Laboratory, College of Materials Innovation and Technology, King Mongkut’s Institute of Technology Ladkrabang, Ladkrabang, Bangkok 10520, Thailand}
\author{Sirichok Jungthawan}
\affiliation{School of Physics, Institute of Science, and Center of Excellence in Advanced Functional Materials, Suranaree University of Technology, Muang, Nakhon Ratchasima, 30000, Thailand}
\author{Adisak Boonchun}
\email{adisak.bo@ku.th}
\affiliation{Department of Physics, Faculty of Science, Kasetsart University, Chatuchak, Bangkok 10900, Thailand}

\title[An \textsf{achemso} demo]
{Two-dimensional Penta-NiPS Sheets: Two Stable Polymorphs}

\abbreviations{IR,NMR,UV}

\begin{document}
\pagebreak
	
	\begin{abstract}
		The discovery of new and stable two-dimensional (2D) materials with exotic properties is essential for technological advancement. Inspired by the recently reported \textit{penta}-PdPSe, we proposed \textit{penta}-NiPS as a new member of the \textit{penta}-2D materials based on first-principles calculations. The \textit{penta}-NiPS monolayer is stable in two polymorphs including the $\alpha$ phase with identical structure as \textit{penta}-PdPSe, and the newly proposed $\beta$ phase with rotated sublayers. Comprehensive analyses indicated that both phases are thermodynamically, dynamically, mechanically, and thermally stable. The \textit{penta}-NiPS is a soft material with 2D Young’s modulus of $E_a$ = 208 Nm$^{-1}$ and $E_b$ = 178 Nm$^{-1}$ for the $\alpha$ phase and $E_a$ = 184 Nm$^{-1}$ and $E_b$ = 140 Nm$^{-1}$ for the $\beta$ phase. Interestingly, the $\alpha$-\textit{penta}-NiPS showed nearly zero Poisson's ratios along the in-plane direction, where its dimensions would be maintained when being extended. For electronic application, we demonstrated that \textit{penta}-NiPS is the wide band gap semiconductors with an indirect band gap of 2.35 eV for $\alpha$ phase, and 2.20 eV for $\beta$ phase. 
	\end{abstract}

	\pagebreak
	
	\section{Introduction}
	Since the discovery of hexacyclic graphene, two-dimensional (2D) materials have been rapidly advanced in terms of production and materials design. An increasing demand on their discovery has been driving the theoretical explorations of stable 2D materials beyond the conventional honeycomb-like structure, including the understanding of composition-structure-properties relationships. Recently, Zhang \textit{et al.}\cite{Zhang2372} theoretically proposed a new pentagonal structure \textit{penta}-graphene with carbon atoms uniquely resembling Cairo pentagonal tiling. Although not yet experimentally synthesized, the properties of the \textit{penta}-graphene  have been widely explored through first-principles calculations. The wide band gap of 3.25 eV and high strength could lead to  potential applications in nanoelectronics and nanomechanical devices. Other prominent examples are \ce{SiC2}, \ce{CN2}, \ce{BN2}, \ce{B2C}, \ce{PdS2}, \ce{AlN2}, \ce{PtN2}, \ce{PdN2}, and \ce{NiS2} \cite{penta-SiC2,CN2,BN2,B2C,cheng2021pentagonal,penta-PdS2,penta-AlN2,penta-NiS2}. In all cases, it is essential that one thoroughly explores various aspects of materials' stability including dynamic, mechanical, and thermal ones.  
	
	The investigation of ternary compounds provides a higher degree of freedom in tuning the properties of \textit{penta}-2D materials \cite{BCN}. The prominent example is \textit{penta}-BCN \cite{BCN} which is thermodynamically, mechanically, and thermally stable according to density functional theory (DFT), in addition to possessing a high piezoelectric response. Later on, the \textit{penta}-\ce{PdPSe} with intrinsic in-plane anisotropy has been experimentally synthesized.\cite{PdPSe} This material is made up from two PdPSe sublayers which are uncommon in nature, leading to unusual electronic and optical properties.\cite{mortazavi2022,bafekry2022two} Nevertheless, \textit{penta}-\ce{PdPSe} is the closely related member of the binary \textit{penta}-\ce{PdSe2} which can be readily synthesized via mechanical exfoliation from bulk crystals.\cite{pdse2} It has been experimentally demonstrated that the \textit{penta}-\ce{PdSe2} based field-effect transistors (FETs) exhibited band gap tunability and high electron mobility.\cite{pdse2} Accordingly, chemical intuition with appropriate atom substitution is a simple yet potential method for theoretically discovering novel materials. 
	
	Herein, we used first-principles calculations to search for a novel member in the ternary pentagonal sheet family. Based on the reported atomic configuration of \textit{penta}-\ce{PdPSe}, Pd was replaced with Ni, and Se was replaced with S in this work. We found that the proposed composition \textit{penta}-NiPS exists in two polymorphs which are thermodynamically, dynamically, mechanically, and thermally stable. The $\alpha$ phase has the structure similar to \textit{penta}-\ce{PdPSe}. Meanwhile, another stable structure, the $\beta$ phase, can be constructed by rotating the sublayers from the $\alpha$ phase. Notably, the $\alpha$ phase exhibited zero Poisson's ratio (ZPR) in the principle $x$ and $y$ directions (i.e., along the sheet), a unique ability to maintain its shape when being compressed or extended. This is rather unusual because most solids exhibit a positive Poisson's ratio (PPR) in the range of 0.2-0.3,\cite{greaves2011poisson} which describes the level of transversally deformation under an axial strain. The electronic and mechanical properties of two phase have also been compared.

	\section{Computational Method}
Our calculations are based on density functional theory (DFT) and the plane-wave  basis  projector  augmented  wave (PAW)\cite{blochl1994projector}  method, as defined in the  Vienna Ab-initio Simulation Package (VASP)\cite{Kresse1996,Kresse1996_2}. The kinetic energy cutoff of 500 eV is used for the plane-wave expansion. Perdew-Burke-Ernzerhof (PBE)\cite{Perdew1996}  functional is imposed for geometry optimization. In order to accurately describe layered material, the long-range van der Waals (vdW) interactions were taken into account using the DFT-D2 functional proposed by Grimme.\cite{DFT-D2} The structures are optimized until total Hellmann-Feynman forces are decreased to 0.02 eV/\AA  and the energy convergence is set to $10^{-6}$ eV. The centered k-point grid within $7 \times 7 \times 1$ is employed for first Brillouin
	zone integration within the Monkhorst-Pack\cite{monkhorst1976special} scheme. A vacuum layer of 16 \AA\ is included along the direction perpendicular to a crystal layer to avoid the interaction between the layer and its neighboring image. A PHONOPY package\cite{phonopy} over $3 \times 3 \times 1$ supercell with 500 eV of energy cutoff and a single k-point at $\Gamma$ is used to calculate phonon spectra to verify the dynamic stability. Ab initio molecular dynamics (AIMD) simulations is employed to confirm thermodynamic stability.  To perform electronic band structure, since DFT within the standard approximation severely underestimates band gaps, we use the Heyd-Scuseria-Ernzerhof (HSE06)\cite{hse03,hse06} hybrid functional to correct not only the band gap but also the band width.
	
	\section{Results and discussion}
	\begin{figure}[h]
		\begin{center}
			\includegraphics[width=15cm]{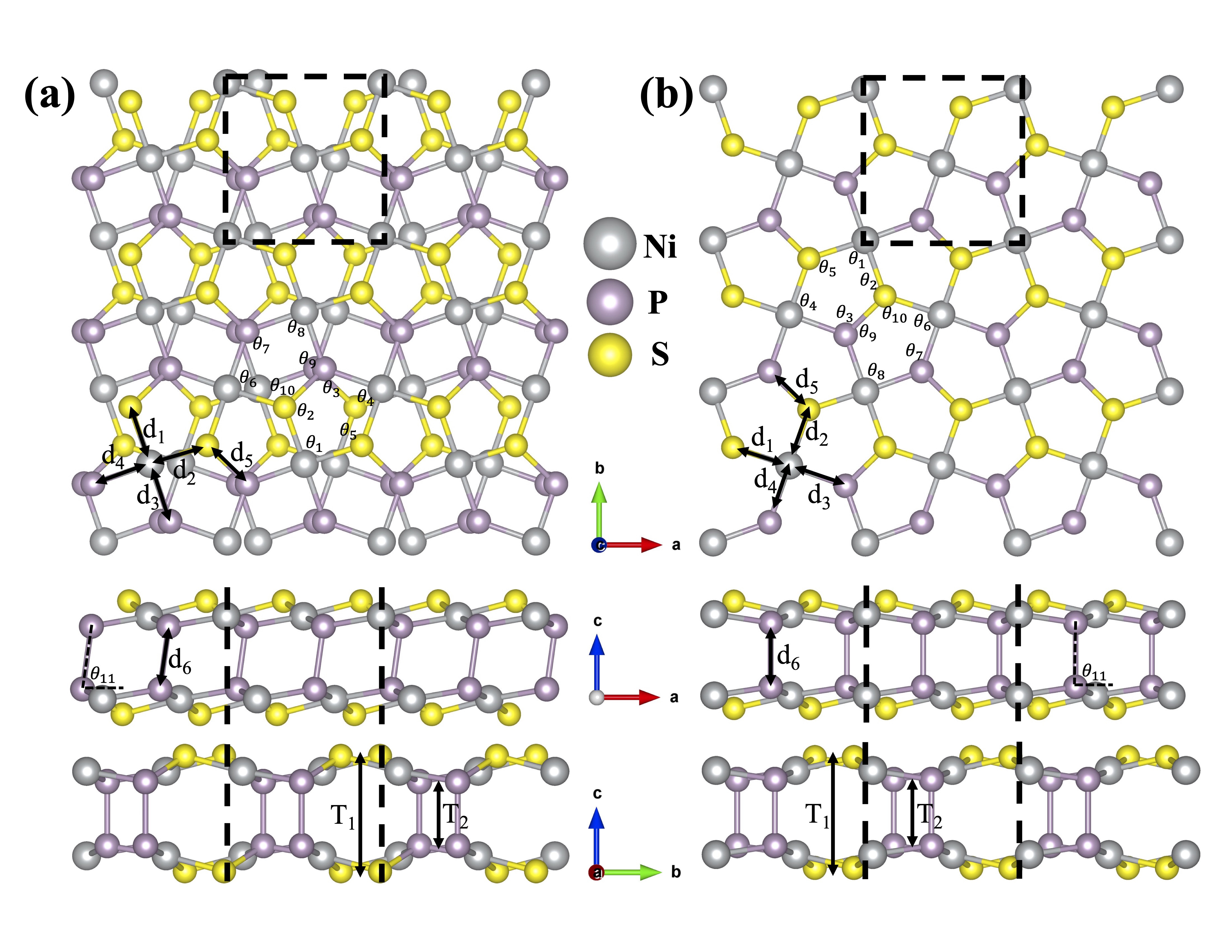} 
			\caption{\ 3x3 supercell of pentagonal (a) $\alpha$-NiPS and (b) $\beta$-NiPS geometrical structures.}
			\label{fig1}
		\end{center}
	\end{figure}
	
	We first constructed the \textit{penta}-NiPS based on the reported atomic configuration of PdPSe, by replacing the Pd atoms with Ni atoms and Se atom with S atoms. This is called the $\alpha$-phase which composes of two sublayers. Within one sublayer, every Ni atom is linked by two P atoms and two S atoms, showing the pentagon structure, as shown in Fig. \ref{fig1} (a). The pentagonal primitive unit cell, shown as the square in Fig. \ref{fig1} (a),  consists of 12 atoms including 4 Ni atoms, 4 P atoms, and 4 S atoms. The lattice constants of the $\alpha$ phase \textit{penta}-NiPS  are \textit{\textbf{a}} = 5.43 \AA\ and \textit{\textbf{b}} = 5.38  \AA. The whole thickness of the layer, T$_1$, is 3.98 \AA, while the distance between two sublayers, T$_2$, is 2.20 \AA\ .  
	
	In addition, we discovered another stable phase, namely the $\beta$ phase, where the two sublayers directly stack on top of each other as shown in Fig. \ref{fig1}(b). The coordinates of the optimized geometry are provided in the Supporting Information (SI). The 12-atom primitive unit cell is shown in the square in Fig. \ref{fig1}(b). 
	The lattice constants of the $\beta$ phase  are \textit{\textbf{a}} = 5.45  \AA\ and \textit{\textbf{b}} = 5.41 \AA, which is slightly larger than $\alpha$ phase. Meanwhile, T$_1$ is 3.86 \AA, and T$_2$ is 2.29 \AA\ . The larger T$_2$ of the $\beta$ phase is understandable because the direct stacking generates a higher coulomb repulsion of the same atoms between two sublayers, as is commonly observed in other 2D materials. The fundamental properties of the two phases including bond lengths, bond angles and band gap are listed in Table. \ref{tbl:lattice}. 
\begin{table}	
	\caption{The calculated lattice constants \textit{\textbf{a}} and \textit{\textbf{b}} in \AA, bond lengths $d_1$-$d_6$ in \AA, bond angles $\theta_1$-$\theta_{11}$ in degree, thicknesses T$_1$ and T$_2$ in \AA, the cohesive energy per atom in eV/atom, and band gap $E_g^{PBE}$ and $E_g^{HSE}$ in eV}
	\label{tbl:lattice}
	\centering\begin{tabular}{llllllllll}
		\hline
		& \textit{\textbf{a}} & \textit{\textbf{b}} & d$_1$ & d$_2$ & d$_3$ & d$_4$ & d$_5$ & d$_6$ &  \\ \hline
		$\alpha$-NiPS & 5.43 & 5.38 & 2.18 & 2.18 & 2.18 & 2.17 & 2.14 & 2.23 &   \\
		$\beta$-NiPS & 5.45 & 5.41 & 2.17 & 2.17 & 2.17 & 2.18 & 2.12 & 2.29 &    \\ \hline
		& $\theta_1$ & $\theta_2$ & $\theta_3$ & $\theta_4$ & $\theta_5$ & $\theta_6$ & $\theta_7$ & $\theta_8$ & $\theta_9$  \\ \hline
		$\alpha$-NiPS & 88.12 & 105.45 & 109.61 & 92.22 & 120.42 & 91.10 & 125.10 & 88.15 & 108.21 \\
		$\beta$-NiPS & 88.71 & 107.72 & 110.31 & 92.18 & 122.06 & 90.63 & 126.21 & 88.21 & 109.66  \\ \hline
		&  $\theta_{10}$ & $\theta_{11}$ & T$_1$ & T$_2$ & $E_{coh}$ & $E_g^{PBE}$ &  $E_g^{HSE}$ & &     \\ \hline
		$\alpha$-NiPS & 108.45 & 83.48 & 3.98 & 2.20 & -5.44 & 0.93 & 2.35 & &   \\
		$\beta$-NiPS & 110.04 & 90.00 & 3.86 & 2.29 & -5.42 & 0.71 & 2.20 & &    \\ \hline
	\end{tabular}
\end{table}

\begin{figure}
		\begin{center}
			\includegraphics[width=10cm]{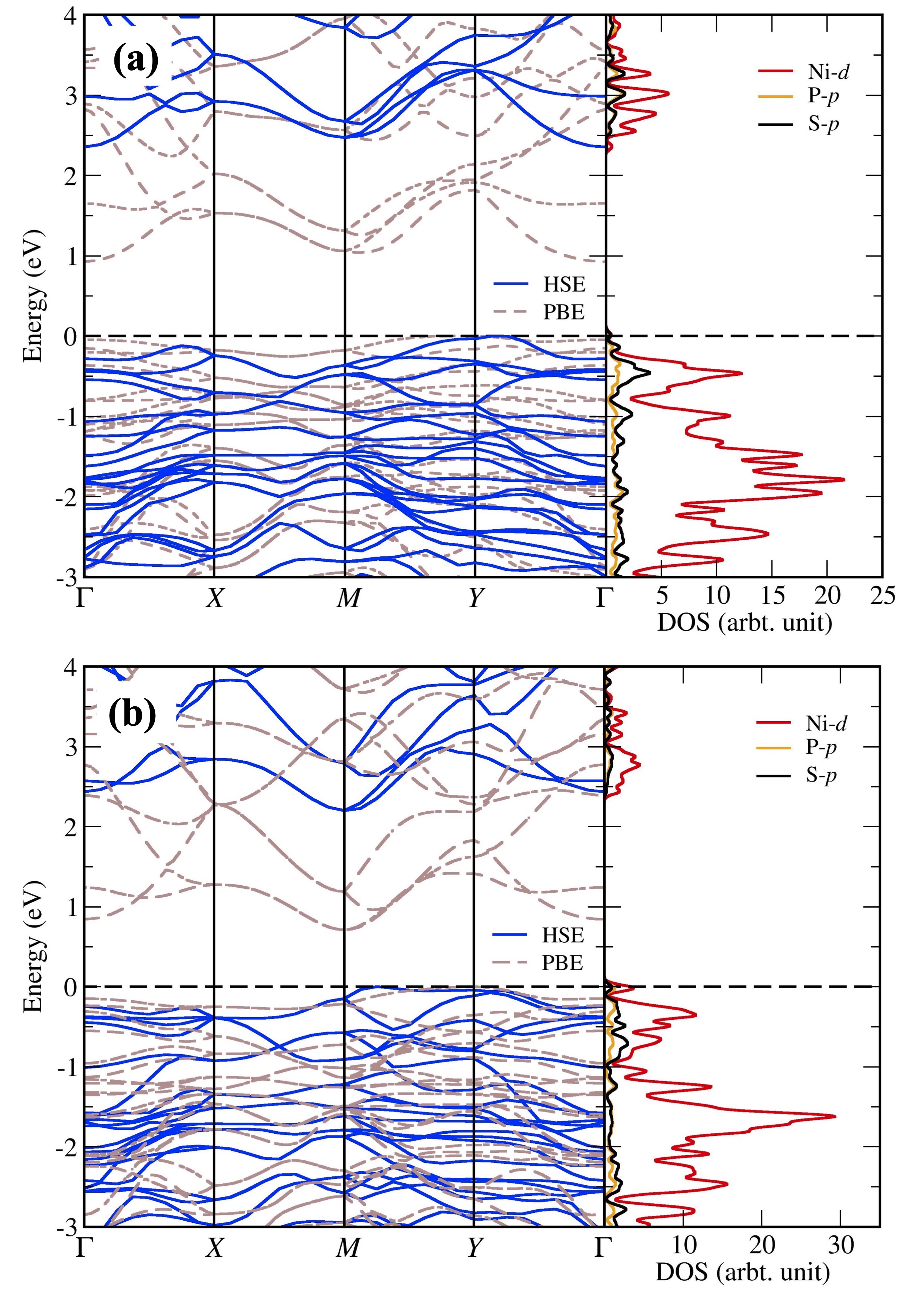} 
			\caption{\ Calculated band structure and partial density of state (PDOS) for  (a) $\alpha$-NiPS and (b) $\beta$-NiPS, respectively.}
			\label{fig:banddos}
		\end{center}
	\end{figure}

	The electronic properties of $\alpha$- and $\beta$-NiPS phases are first elucidated by the calculated band structure as shown in Fig.\ref{fig:banddos}. The dash and solid lines are energy bands calculated from PBE and HSE06 functional, respectively. By using the PBE functional,\cite{Perdew1996} the electronic band structures show indirect band gaps of 0.88 and 0.68 eV for $\alpha$ and $\beta$ phase, respectively. As it is well-known that the standard DFT underestimates band gap, the HSE06 functional\cite{hse03,hse06} was also employed. It is found that a larger band gap of 2.35 eV was obtained for the $\alpha$ phase, vs 2.20 eV for the $\beta$ phase, with the indirect band gap character being consistent. As seen in  Fig.\ref{fig:banddos} within HSE06, the valence band maximum (VBM) of $\alpha$ phase is located along the $M-Y$ path and the conduction band minimum (CBM) at the $M$ point. In the $\beta$ phase, the VBM is located along the $\Gamma-Y$ path and the CBM is at the $\Gamma$ point. 
	
	In order to analyze the electronic properties of \textit{penta}-NiPS, the partial density of states (PDOS) obtained from the HSE06\cite{hse03,hse06} functional were plotted along with the band structure in Fig.\ref{fig:banddos}. For $\alpha$-NiPS, the VBM is dominated by the P-$p$ and the S-$p$ state, while the CBM has a main contribution from the Ni-$d$ states. For $\beta$-NiPS, both of VBM and CBM are mainly contributed by the Ni-$d$ states. The average $d$ band center of Ni in the $\alpha$ phase is $\sim1.85 \, {\rm eV}$ with respect to VBM, while the average $d$ band center in the $\beta$ phase is $\sim1.65\, {\rm eV}$ with respect to VBM. It is known that the deeper energy of the $d$ states weakens the $p-d$ coupling, leading to a further increase in band gap \cite{PhysRevB.74.045202} as observed in the present work.

	To evaluate the crystal binding stability, the cohesive energy per atom was calculated, given by the following equation:
	\begin{equation}
		E_{coh} = \frac{E_{tot}(\text{NiPS})-\sum n_i E_i}{\sum n_i}
	\end{equation}
	where $E_{tot}(\text{NiPS})$ is the total energy of the \textit{penta}-NiPS, $n_i$ is the number of Ni, P and S atoms in the unit cell and $E_i$ represents the energies of an isolated single Ni, P and S atom. The calculated cohesive energy of the \textit{penta}-NiPS are -5.44 and -5.42 eV/atom for the $\alpha$ and $\beta$ phase, respectively. The negative values indicate that the \textit{penta}-NiPS are more energetically favorable than the isolated atoms. In addition, the $\alpha$ phase is just slighlty more energetically favorable than the $\beta$ phase. Generally, different polymorphs (e.g., anatase vs rutile for $\rm TiO_2$) could be synthesized by a fine tuning of synthetic conditions such as temperature, pressure, surrounding gas, among others.\cite{wells2012structural} The selective synthesis of either $\alpha$ and $\beta$ \textit{penta}-NiPS remains to be accomplished.
	
		\begin{figure}[h]
		\begin{center}
			\includegraphics[width=10cm]{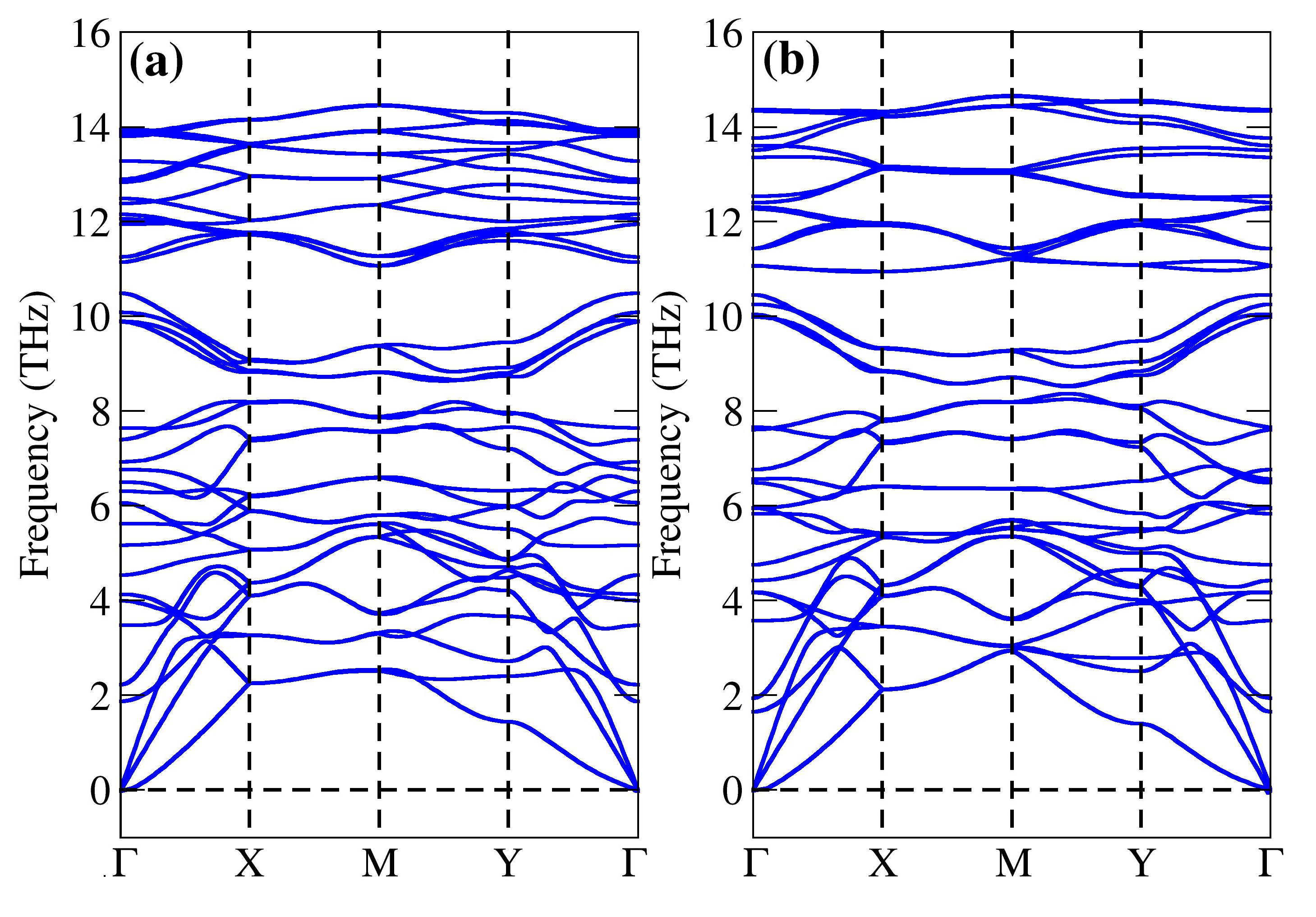} 
			\caption{\ Calculated phonon band structure of (a) $\alpha$-NiPS and (b) $\beta$-NiPS.}
			\label{fig2}
		\end{center}
	\end{figure}

We verify the dynamic stability of $\alpha$ and $\beta$ \textit{penta}-NiPS by calculating their phonon dispersion. The dynamic stability manifests itself as the positive frequencies in the phonon spectra throughout the entire  Brillouin zone. On the other hand, if the spectra display imaginary frequency modes (negative values), the structures reveal non-restorative force against a displacement of atoms, indicating instability. Fig \ref{fig2}(a) and Fig \ref{fig2}(b) show the calculated phonon dispersion of \textit{penta}-NiPS for $\alpha$ and $\beta$ phase, respectively. There are no imaginary modes throughout the Brillouin zone, confirming that both are dynamically stable. We note that the imperceptible negative frequencies at $\Gamma$, (visible only when zoomed in) is a well-known computational error which can be effectively eliminated using the LDA functional, for example. These negative frequencies of phonon dispersion are also observed in \textit{penta}-\ce{PdSe2}.\cite{pdse2}
	
	Mechanical stability is an important property for strain engineering. Such stability can be verified from the linear elastic constants, C$_{11}$, C$_{22}$, C$_{12}$ and C$_{66}$, which are elements in the stiffness tensor. Here, the elastic constants are obtained by finite difference method. Our calculated elastic constants of both polymorphs are listed in table 2. Typically, the structures are considered mechanically stable when the linear elastic constants satisfy the conditions C$_{11}$C$_{22}$-C$_{12}^2 > 0$ and C$_{66} > 0$. Notably, these conditions are obeyed for both phases, confirming the mechanical stability of \textit{penta}-NiPS.
	
	\begin{table}
	\caption{The elastic constants $C_{jk}$ (Nm$^{-1}$), Poisson's ratio on [100] ($\nu_a$) and [010] ($\nu_b$) directions and Young's modulus (Nm$^{-1}$) on [100] ($E_a$) and [010] ($E_b$) directions.}
	\label{tbl:constant}
	\begin{tabular}{lll}
		\hline
		& $\alpha$-NiPS & $\beta$-NiPS \\ \hline
		$C_{11}$  & 184.27 & 211.55 \\
		$C_{12}$ & 0.52 & 28.52 \\
		$C_{22}$  & 140.50 & 182.60 \\
		$C_{66}$  & 65.03 & 66.48 \\
		$\nu_a$ & 0.003 & 0.135 \\
		$\nu_b$ & 0.004 & 0.156 \\
		$E_a$  & 184.27 & 207.71 \\
		$E_b$ & 140.50 & 178.14 \\
		\hline
	\end{tabular}
\end{table}

	Next, we evaluated the 2D Young’s modulus on the [100] and [010] directions (i.e., along the plane), which is obtained by $E_a =$(C$^2_{11}$ - C$^2_{12}$)/C$_{11}$ and $E_b =$ (C$^2_{22}$ - C$^2_{12}$)/C$_{22}$. In the $\alpha$ phase, $E_a$ = 207.71 Nm$^{-1}$ and $E_b$ = 178.14 Nm$^{-1}$. These values are close to those of \textit{penta}-PdPSe with identical structure, $E_a$ = 218 Nm$^{-1}$ and $E_b$ = 152 Nm$^{-1}$, calculated from DFT calculation.\cite{mortazavi2022} Therefore, the $\alpha$-NiPS are as strong as \textit{penta}-PdPSe. In $\beta$ phase, the calculated values are $E_a$ = 184.27 Nm$^{-1}$ and $E_b$ = 140.50 Nm$^{-1}$, indicating that it is softer than the $\alpha$ phase. Table \ref{tbl:constant} lists the calculated Poisson’s ratio. For $\beta$-NiPS, the calculated Poisson’s ratio on the [100] direction, $\nu_a$ = C$_{12}$/C$_{11}$ is 0.135 which is slightly lower than that on the the [010] direction, $\nu_b$ = C$_{12}$/C$_{22}$ of 0.156. Interestingly, for $\alpha$-NiPS, the calculated Poisson’s ratios are nearly zero for both [100] and [010]  directions ($0^{\circ}$ and $90^{\circ}$), $\nu_a$ = 0.003 and $\nu_b$ = 0.004. The nearly zero Poisson’s ratios was previously observed in ternary penta-BCN bus it is located near [110] direction ($\sim45^{\circ}$) with the value of 0.004.\cite{BCN} Moreover, Hou \textit{et al.} \cite{bcp} reported that the ternary \textit{penta}-BCP exhibited the largest positive Poisson’s ratio of 1.30 along the [010] direction, which is now the highest among 2D materials. The near zero Poisson’s ratios of the present $\beta$-NiPS indicate that $\beta$-NiPS has an ability to maintain its dimensions under compression/extension along the in-plane directions.  
	
\begin{figure}[h]
	\begin{center}
		\includegraphics[width=9.5cm]{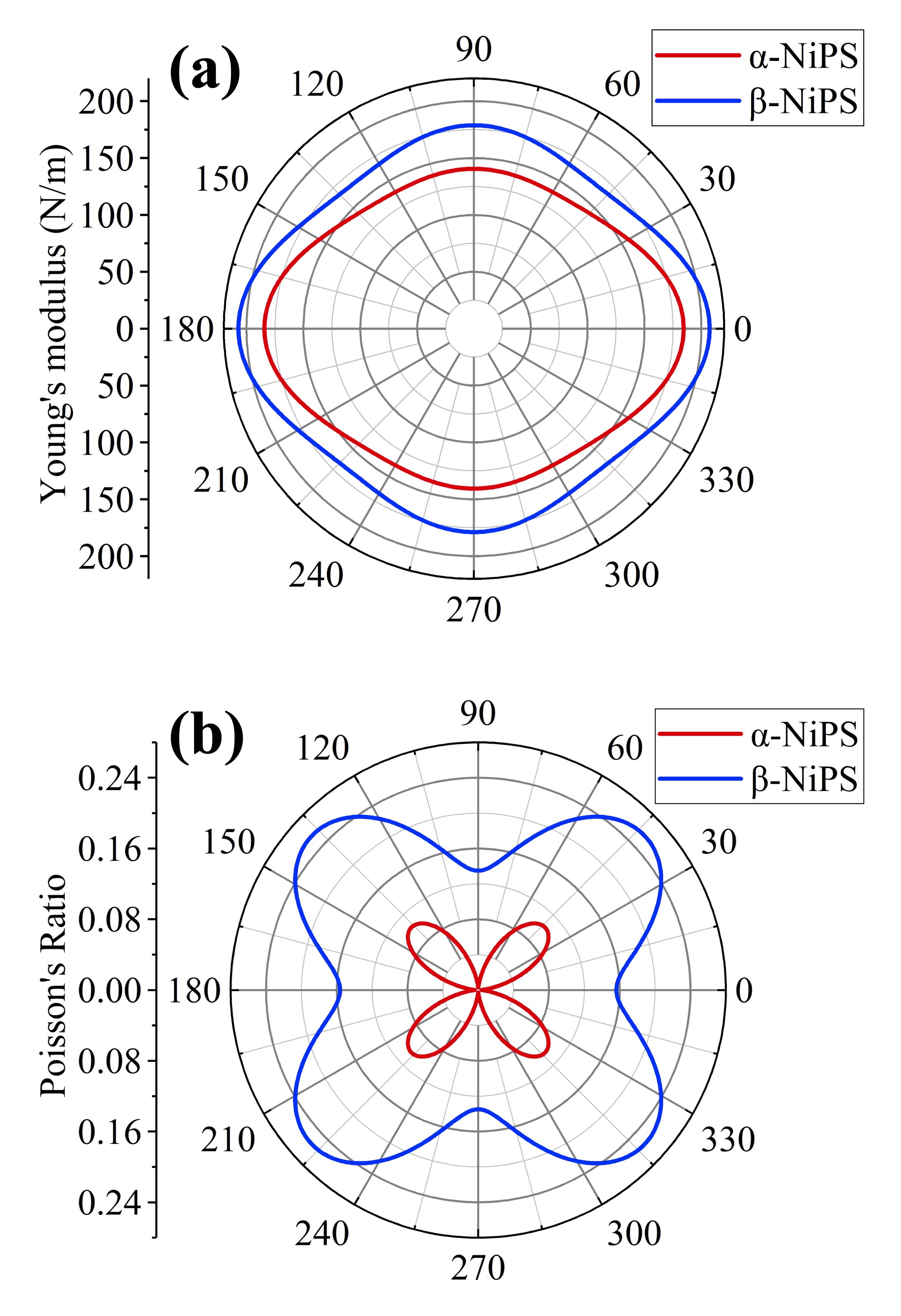} 
		\caption{\  Variation of (a) in-plane Young's modulus,  and (b) Poisson's ratio for $\alpha$-NiPS and $\beta$-NiPS. Note that solid  and dash line represent $\alpha$-NiPS and $\beta$-NiPS, respectively.}
		\label{fgr:polar}
	\end{center}
\end{figure}
	
	To further explore the dependence of mechanical properties on the crystal orientation of \textit{penta}-NiPS, we computed the Young's modulus ($E$) and Poisson's ratio ($\nu$) along an arbitrary direction $\theta$ ($\theta$ is the angle relative to the x direction) using the following formulas:
	
	\begin{eqnarray}
		E(\theta) = \frac{C_{11}C_{12}-C_{12}^2}{C_{11}s^4+C_{22}c^4+\left(\frac{C_{11}C_{12}-C_{12}^2}{C_{66}}-2C_{12}\right)c^2 s^2} \\ [3pt]
		\nu(\theta) = \frac{\left(C_{11}+C_{12}-\frac{C_{11}C_{12}-C_{12}^2}{C_{66}}\right)c^2 s^2-C_{12}(c^4+s^4)} {C_{11}s^4\theta+C_{22}c^4+\left(\frac{C_{11}C_{12}-C_{12}^2}{C_{66}}-2C_{12}\right)c^2 s^2} 
	\end{eqnarray}
	where $c=\cos\theta$ and $s=\sin\theta$. The calculated results are plotted in Fig \ref{fgr:polar}. One can see that the maxima of the in-plane Young’s modulus in both phases are located at the [100] direction with the value of 184.27 ($\alpha$-NiPS) and 207.71 ($\beta$-NiPS) ${\rm Nm^{-1}}$, respectively. Meanwhile, the minima are along the [010] direction (90$^{\circ}$) with the value of 140.50 ($\alpha$-NiPS) and 178.14 ($\beta$-NiPS) ${\rm Nm^{-1}}$. In $\alpha$-NiPS, the minimum Poisson’s ratios (which is very close to zero) are located along the [100]  ($0^{\circ}$), while the maximum is along  [110] direction (43$^{\circ}$) with values of 0.10.  In $\beta$-NiPS, the minimum and maximum Poisson’s Ratios are located with the same angle as $\alpha$ phase but at higher positive values. Because the in-plane Young’s modulus and Poisson’s ratios are dependent on crystal orientation, \textit{penta}-NiPS exhibit anisotropic mechanical properties which can be explained by its lattice constant where $a\neq b$.

		\begin{figure}[h]
		\begin{center}
			\includegraphics[width=12cm]{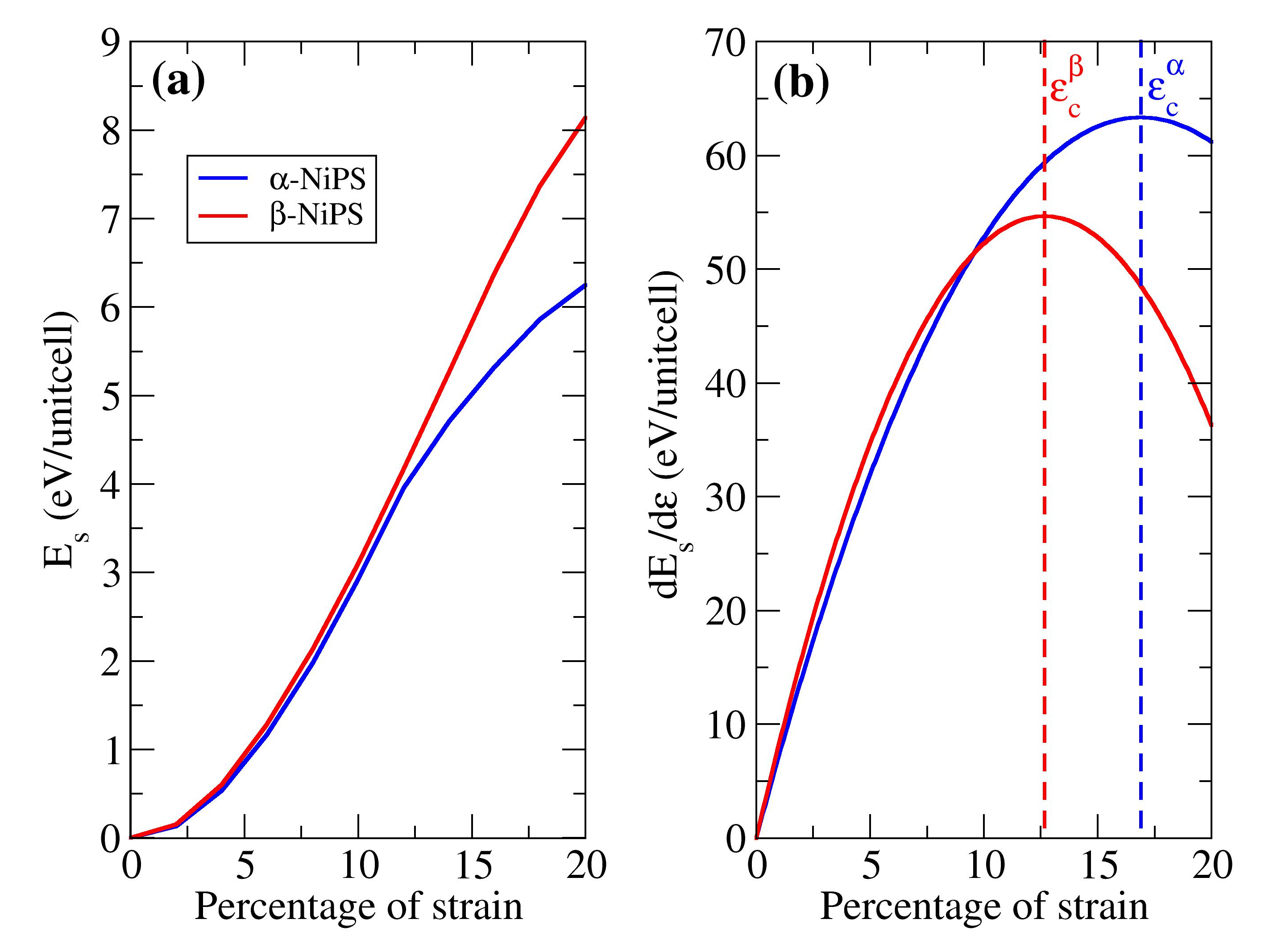} 
			\caption{\ (a) the variation in biaxial strain energy (E$_s$) and (b) its derivative for  $\alpha$-NiPS and $\beta$-NiPS. Blue and red color represent $\alpha$-NiPS and $\beta$-NiPS, respectively.}
			\label{fig4}
		\end{center}
	\end{figure}
	
	For practical application of semiconductors, it is beneficial that one can modify the electronic properties by a simple method such as strain engineering. This can be effectively applied to 2D materials via lattice mismatch on the substrate, thermal expansion, or mechanical loading. Because 2D materials tend to fold up under lateral compression, we focused on the tensile strain only, specifically the biaxial strain effect. The degree of strain is defined as $\varepsilon = \Delta l/l_0 \times 100$, where $\Delta l$ is the difference in a strained length, $l$, and an original, unstrained length, $l_0$. Fig.\ref{fig4} is the plot showing the strain energies, $E_S$, and their derivative ($dE_S(\varepsilon)/d \varepsilon$) with respect to the applied strains from  0 - 20\%. For simplicity, the energies of the strain-free structures are set to zero so that the strain energies, $E_S$ is the energy differences between the total energies of the structure with and without strain. When the biaxial strain is continuously applied, a crystal expands until reaching the mechanical failure at $\varepsilon=\varepsilon_{C}$ where the maximum derivative of strain is observed. The mechanical failure of the $\alpha$-NiPS and $\beta$-NiPS occurs at $\varepsilon_{C}^{\alpha}=$ 16.90\% and 12.66\%, respectively. These values can be compared with 24.3\% for \textit{penta}-BCN, and 17.5\% for hydrogenated \textit{penta}-BCN (H-BCN).\cite{dabsamut2021strain,hbcn} Normally, 2D materials can sustain much larger strains than their bulk structure. Also, 2D materials with buckling structure can withstand a higher strain than the non-buckling structure. Here, the resilience of the $\alpha$ phase could be linked to the bucking height $\rm T_1$ compared to that of the $\beta$ phase. 
	
	It is known that the band gaps of 2D materials can be potentially modulated by strain. Thus, we also investigated how the band gaps of $\alpha$-NiPS and $\beta$-NiPS change with the additional strain, starting from their equilibrium (i.e., zero strain) up to just prior to the critical point of strain.  In Fig. S1(a) and S1(b) in the supplementary information (SI), the band gaps of both polymorphs under the strain seem to linearly decrease from 2.35 to 0.94 eV ($\alpha$) and 2.20 to 1.00 eV ($\beta$), respectively. In $\alpha$-NiPS, the valence band maximum (VBM) shifted from $Y-\Gamma$ path to $Y-M$ path while that of the conduction band minimum (CBM) changed from $M$ point to $\Gamma$ point. On the other hand, the characteristics of VBM and CBM of $\beta$-NiPS have not been altered with strain as depicted in Fig. S2(a) and S2(b) in SI.

	\begin{figure}[h]
	\begin{center}
		\includegraphics[width=15cm]{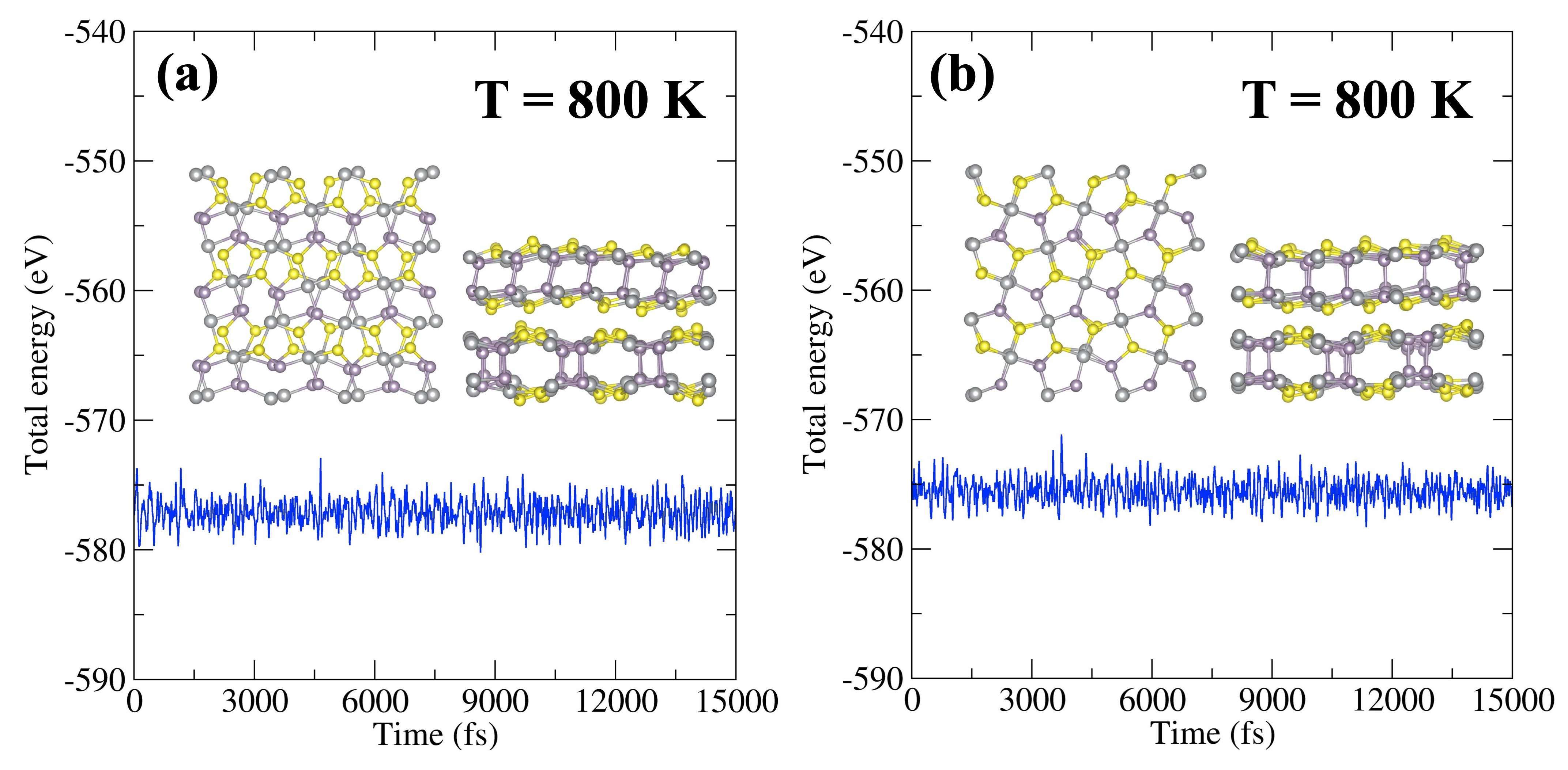} 
		\caption{\ Total energy fluctuation with time during the AIMD simulation at 800 K of (a) $\alpha$-NiPS and (b) $\beta$-NiPS.}
		\label{fig3}
	\end{center}
\end{figure}
	
	The thermal stability of \textit{penta}-NiPS was examined via ab initio molecular dynamic simulations. Supercells with a duplication of a $4 \times 4 \times 1$ of the unit cell is used. The molecular dynamics simulations were performed for 15 ps with a time step of 1 fs under the temperature from 300 to 800 K. Fig. \ref{fig3} shows the fluctuation of the potential energy of \textit{penta}-NiPS throughout the simulation time at 800 K. For both phases, the respective structure reaches the steady state and equilibrates around a constant energy after 1,000 fs. These results clearly indicate the thermal stability of \textit{penta}-NiPS at least up to 800 K.
	
	It is true that both polymorphs of NiPS has not been experimentally realized. Still, the related composition MPS$_3$ (M = Mn, Fe, Co, Ni and Cd) is a classic example of layered compounds crystallizing in CdCl$_2$-type structure.\cite{CdCl2} Although the two structures ($\alpha$- and $\beta$-\textit{penta}-NiPS vs CdCl$_2$) are clearly different, NiPS might be experimentally synthesized. Consider, for example, PdSe$_2$ which can be successfully synthesized by a self-flux method.\cite{pdse2} Moreover, very recently, penta-PdPS (the same structure with $\alpha$-NiPS) can be successfully synthesized by exfoliated from bulk crystals via chemical vapor transport (CVT) method.\cite{PdPS} It is premature to explicitly suggest the synthetic procedure at present, but we hope it could be achievable.

	\subsection{Conclusions}
	By using DFT calculation, we theoretically discovered \textit{penta}-NiPS as the new member in the \textit{penta}-2D family. The \textit{penta}-NiPS is stable in two polymorphs. The $\alpha$-NiPS has the same structure as the previously reported \textit{penta}-PdPSe, while the $\beta$-NiPS was obtained by rotating the two sublayers from the $\alpha$-NiPS. Both phases are  dynamically, mechanically, and thermally stable, as comprehensively verified from  phonon dispersion, elastic constants, and molecular dynamic simulation, respectively. We found that the \textit{penta}-NiPS is a semiconductor with an indirect band gap of 2.35 and 2.20 eV, for $\alpha$ phase and $\beta$ phase. They are soft materials with 2D Young’s modulus of $E_a$ = 207.71 Nm$^{-1}$ and $E_b$ = 178.14 Nm$^{-1}$ for the $\alpha$ phase and $E_a$ = 184.27 Nm$^{-1}$ and $E_b$ = 140.50 Nm$^{-1}$ for the $\beta$ phase. Interestingly, the $\alpha$ \textit{penta}-NiPS exhibited a close to zero Poisson’s ratio of 0.003-0.004. These values are significantly smaller than those of the $\beta$ phase, 0.135-0.156. The nearly zero Poisson’s ratios allow $\beta$-\textit{penta}-NiPS to maintain its dimensions under extension with potential applications in wearable electronics.

	\begin{acknowledgement}
	This project is funded by National Research Council of Thailand (NRCT) and Kasetsart University : N42A650278. K.D. was supported by the National Research Council of Thailand (NRCT), No. NRCT5-RGJ63002-028. T.T. was supported by Graduate Program Scholarship from the Graduate School, Kasetsart University.  S.J. has received funding support from the NSRF via the Program Management Unit for Human Resources \& Institutional Development, Research and Innovation [grant number B05F640051]. We wish to thank NSTDA Supercomputer Center (ThaiSC) for providing computing resources for this work. 
	\end{acknowledgement}

	\begin{suppinfo}
	The coordination of the optimized geometries of $\alpha$ and $\beta$ \textit{penta}-NiPS and additional figures as mentioned in the text.
\end{suppinfo}
	
	\bibliography{penta-NiPS}
	
\end{document}